\documentclass[review]{elsarticle}
 
\usepackage{lineno,hyperref}
\modulolinenumbers[5]
\journal{Nuclear Instruments and Methods in Physics Research}

\usepackage[version=4]{mhchem}
\usepackage{amsmath, amsfonts, amssymb, amsthm, siunitx}

\usepackage{url}
\makeatletter
\g@addto@macro{\UrlBreaks}{\UrlOrds}
\makeatother

\usepackage[section]{placeins}

\begin{document}

\begin{frontmatter}

%\title{Elsevier \LaTeX\ template\tnoteref{mytitlenote}}
\title{Estimating Particle Size for Therapeutic Application of Boron in Proton Therapy using the Finite Element Method}

\author{Jacob D. Baxley \corref{mycorrespondingauthor}}
\cortext[mycorrespondingauthor]{Corresponding author}
\ead{jacob.baxley351@topper.wku.edu}

\author{Duncan Weathers \corref{cor3}}

\author{Tilo Reinert \corref{cor2}}

\address{Ion Beam Modification and Analysis Laboratory, Physics Department, University of North Texas, 1155 Union Circle, No. 311427,
Denton, TX 76203, USA}




\begin{abstract}
Previous measurements have shown large cross sections for the \ce{^{11}B(p,\alpha)^{8}Be} reaction and K-shell ionization of boron from \ce{H^+} ions.  Past publications have shown that this reaction will likely increase the efficacy of proton therapy.  This study investigates the size of boron particles for optimum treatment enhancement in proton therapy.  Simulations of protons passing through varying sized boron particles were developed to compare energy outputs for alpha particles and low energy electrons.  The results for the boron particle radius that produced the largest radiation output are presented in graphical form in this paper.  The radius that produced the largest Auger output was determined to be \SI{1.3}{\nano \meter}.  The results indicate that maximum dose enhancement will depend on the limiting factors of the biological system in regards to the appropriately sized particle.  Studying different reactions that may be applied in hadron therapies allows researchers and physicians to target tumor sites more selectively.

\end{abstract}


\end{frontmatter}

%\linenumbers

\section{Introduction}

\begin{figure}[hbt]
  \includegraphics[width=1.0\textwidth]{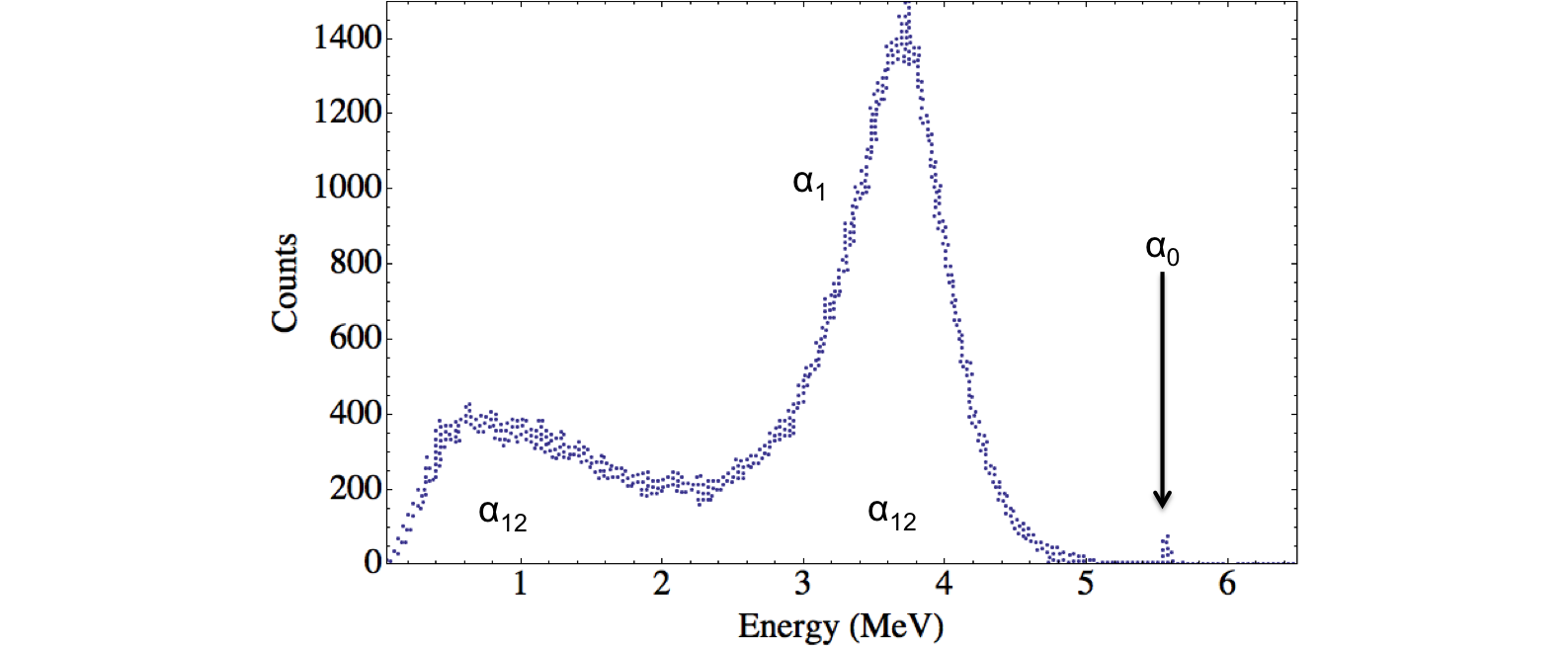}
  \caption{Energy spectrum of \ce{\alpha}-particle yield from the \ce{^11B(p,\alpha)^{8}Be} reaction using a proton beam energy of 660 keV with a detector at 150$^\circ$ to the incident proton direction \cite{liuCrosssections11B8Be2002}.}
  \label{fig:alphaspectnoE}
\end{figure}

Over the last few decades there has been increased interest in further improving radiation therapies.  More than 50\% of patients with localized malignant tumors undergo radiation treatment as a component of their medial care \cite{loefflerChargedParticleTherapy2013}.  A critical challenge in radiation oncology is how to better target tumor volumes while minimizing risks to normal healthy tissue.  One method involves delivering an element or compound to the area of interest then utilizing nuclear reactions for dose enhancement.  This process has been used in boron neutron capture therapy (BNCT) in which $\ce{^10B}$ is irradiated with low energy thermal neutrons to yield high linear energy transfer (LET) alpha particles and $\ce{^7Li}$ nuclei \cite{miyatakeBoronNeutronCapture2016, barthBoronNeutronCapture2005}.  More recently there has been growing interest in the use of platinum and gold nanoparticles in radiotherapies \cite{porcelPlatinumNanoparticlesPromising2010, hainfeldUseGoldNanoparticles2004}.  The nanoparticles can act as radiosensitizers by producing low energy electrons and x rays upon ionization that cause single and double DNA strand breaks in a target volume \cite{championQuantummechanicalPredictionsElectroninduced2013}.  The electrons also induce water radiolysis and production of $^{\circ}$OH hydroxyl radical clusters.  These free radicals produce further breaks in the DNA \cite{usamiComparisonDNABreaks2010}.  Due to the limited range of the electrons the Auger electron cascade may benefit ion beam radiotherapies by enhancing the dose in targeted regions.  %Previous studies have investigated the use of gold nanoparticles in proton therapy \cite{polf2011enhanced} and molecules containing platinum with other hadrontherapies \cite{usami2007irradiation}.  Other metal nanoparticles have been simulated with proton interaction \cite{walzlein2014simulations}.
Nanoparticles may be advantageous in creating secondary electrons arising out of Auger cascades utilizing high Z (atomic number) atom excitation.  However, dense high Z materials also possess a high stopping power for low-energy electrons.  This high stopping power may hinder electrons from escaping the nanoparticles to the targeted tissue and thus limit their effectiveness.  Boron (z = 5) is a metalloid with a high Auger electron yield per electron from KLL ionization.  Boron delivery agents are already being studied in BNCT \cite{barthBoronNeutronCapture2005, miyatakeBoronNeutronCapture2016}.  While these characteristics  show promise in enhancing radiotherapy, it must be considered that the boron, in nanoparticle form, may absorb a significant amount of electron energy, effectively ``self-shielding" the site from treatment.
In addition to the Auger effect, boron particles may also undergo the \ce{^11B(p,\alpha)^{8}Be} reaction \cite{meyerTechnicalNoteMonte2022}.  Previous studies show that proton boron fusion therapy(PBFT) enhances dose in targeted tissue \cite{cirroneFirstExperimentalProof2018, kimSimulationStudyRadiation2017}.  This reaction produces three alpha particles through the two channels shown here and in Figure \ref{fig:alphaspectnoE}:
\begin{equation}
\label{eq:a0}
  \ce{^11B {+} p -> \alpha_0 {+} ^8{Be} {+} (Q=8.586 MeV)}
\end{equation}

\begin{equation}
\label{eq:a1}
  \ce{^11B {+} p -> \alpha_1 {+} ^8{Be}^* {+} (Q=5.65 MeV)}
\end{equation}

\begin{equation}
\label{eq:Be}
  \ce{^8Be -> \alpha_12 {+} \alpha_12 {+} (Q=0.092 MeV)}
\end{equation}

\begin{equation}
\label{eq:Bes}
  \ce{^8Be^* -> \alpha_12 {+} \alpha_12 {+} (Q=3.028 MeV)}.
\end{equation}

 Alpha particles have ten times the radiation weighting factor in tissue of protons and twenty times the radiation weighting factor of photons \cite{wrixonNewICRPRecommendations2008}.  The proton-boron reaction is exothermic with a Q value totaling 9.506 MeV. A large cross section is found at a broad resonance peak at $\approx$660 keV.  These characteristics make this reaction attractive for enhancing proton therapy. \cite{liuCrosssections11B8Be2002}

Another type of therapy, embolization therapy, involves the introduction of small particles into tissue to reduce the blood flow to elicit vascular blockade and promote tumor necrosis.  Embolization has also been used with drugs or chemotherapy, in which case it is known as chemoembolisation.  Benefits from embolization therapies are dependent on blood supply, which limits the types of tissue treated.  Embolic agents have been primarily investigated in treating liver cancers, though the method shows promise in the treatment of pancreatic cancer and breast cancer  \cite{taguchiChemoOcclusionTreatmentLiver1994}.  Embolization therapy could in principle be used in combination with PBFT to treat these cancers by using \ce{^{11}B} microparticles.  The efficacy of this approach would depend on what sized boron particle yields the greatest radiation dose to the surrounding tissue.

The aim of this study is to simulate and compare this useful radiation energy output for differing sized boron particles under proton bombardment.  This simulation will allow us to better assess what sized boron particle will produce the greatest dose enhancement in proton therapy.  These results will assist medical physicists in selecting a boron particle size to study for clinical analysis.

\section{Methods}

Mathematica \cite{wolframresearchincMathematica2019} provided the environment used to program the \ce{^11B(p,\alpha)^8Be} and Auger simulations for varying boron particle sizes. These codes can be found on GitHub at \cite{baxleyPBFTParticleSize}.  The objective of each simulation was to find the size of spherical boron particle resulting in the greatest amount of energy deposition to the surrounding medium.  The energy of each escaping alpha particle was totaled into a summation termed ``alpha energy product."  Similarly, the total escaping electron energy was termed ``electron energy product."  Each simulation used a two-dimensional polar scattering geometry with results suitably weighted to translate to three dimensions.

\begin{figure}[hbt]
  \includegraphics[width=1.0\textwidth]{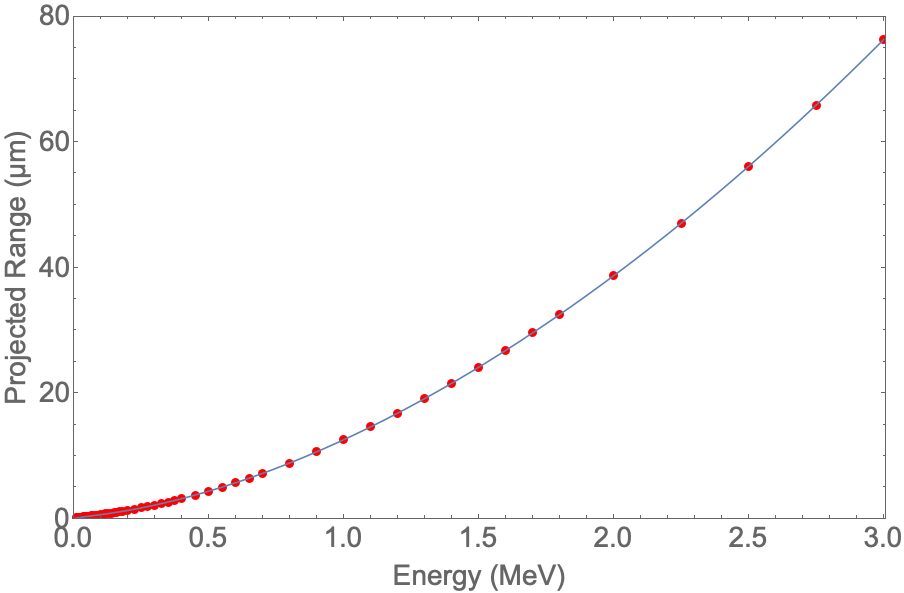}
  \caption{Projected range for protons in boron as a function of proton energy.  Data from SRIM (circles) \cite{zieglerSRIMStoppingRange2008} are shown, along with the fit from Mathematica's NonlinearModelFit function (line) \cite{wolframresearchincMathematica2019}.}
  \label{fig:PRinB}
\end{figure}

\begin{figure}[hbt]
  \includegraphics[width=1.0\textwidth]{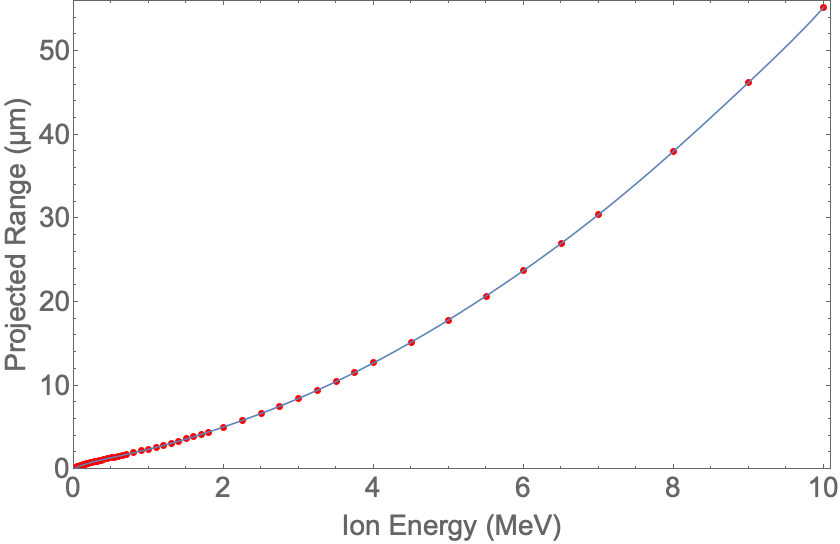}
  \caption{Projected range for alpha particles in boron as a function of alpha particle energy.  Data from SRIM (circles) \cite{zieglerSRIMStoppingRange2008} are shown, along with the fit from Mathematica's NonlinearModelFit function (line) \cite{wolframresearchincMathematica2019}.}
  \label{fig:PRHeinB}
\end{figure}

\begin{figure}[hbt]
  \includegraphics[width=1.0\textwidth]{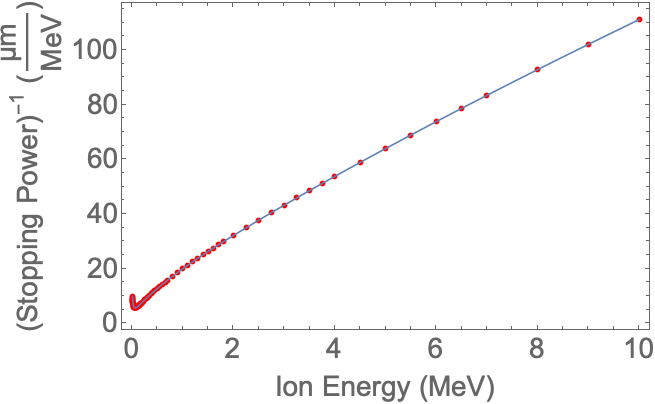}
  \caption{Reciprocal of stopping power for protons in boron as a function of proton energy.  Data from SRIM (circles) \cite{zieglerSRIMStoppingRange2008} are shown, along with the fit from Mathematica's NonlinearModelFit function (line) \cite{wolframresearchincMathematica2019}.}
  \label{fig:RecipStoppingPower}
\end{figure}

\begin{figure}[hbt]
  \includegraphics[width=1.0\textwidth]{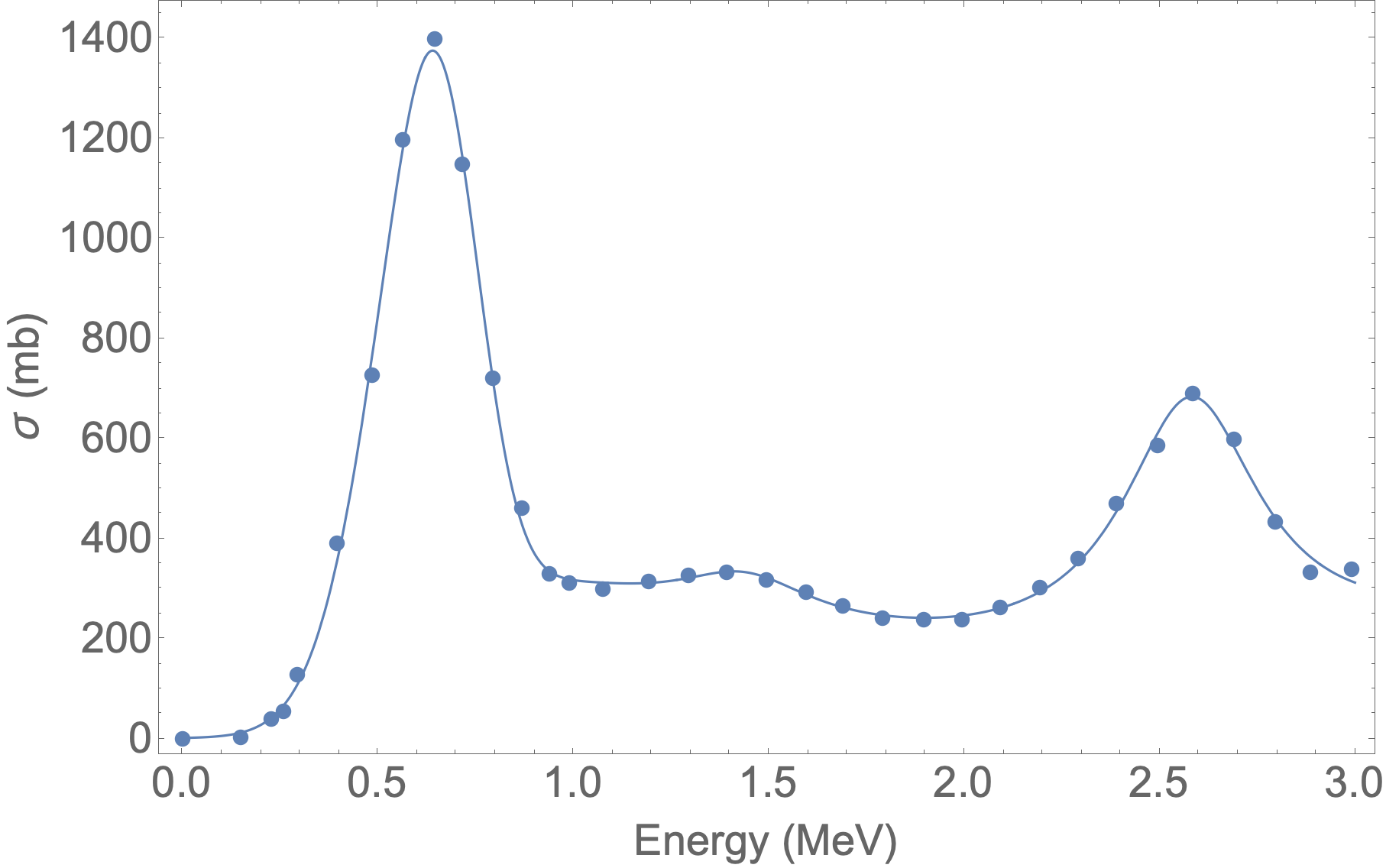}
  \caption{Total cross section $\sigma$ of the \ce{^11B(p,\alpha)^8Be} reaction (circle). Theses data have been scaled to account for the multiple alphas. Data from Sikora \cite{sikoraNewEvaluation11B2016} were evaluated using Mathematica's NonlinearModelFit function (line) \cite{wolframresearchincMathematica2019}.}
  \label{fig:sikora}
\end{figure}

SRIM (The Stopping and Range of Ions in Matter) was used for both proton and alpha particle energy loss estimations and stopping power calculations in all materials \cite{zieglerSRIMStoppingRange2008}.  Datasets for projected ranges, stopping powers, cross sections and related quantities were modeled with Mathematica's NonlinearModelFit function.  The results of these fits for the proton-boron reaction are shown in Figure \ref{fig:PRinB} through Figure \ref{fig:sikora}.

The \ce{^11B(p,\alpha)^8Be} reaction's angular dependent differential cross section, $\frac{d \sigma}{d \Omega}$, was provided by Sikora \cite{sikoraNewEvaluation11B2016}.  Angular dependence of the differential cross sections were fitted with a Legendre polynomial expansion in the center of mass frame and scaled.  Coefficients based on energy were provided in tabulated form. Cross section can have a significant effect on reaction rates and therefore simulation results.  Previous existing cross section data can have errors up to 30\% and inconsistencies as high as 50\% \cite{liuCrosssections11B8Be2002}. More than 95\% of Sikora's individual data points were within 3\% of the fitted results.

%The unrestricted LET was calculated using the electronic stopping power from SRIM.  The LET of each alpha was integrated to determine the absorbed dose.  The same was done to the unrestricted LET for protons.  From this the ratio for equivalent doses were compared for the same mass of water. \cite{Seltzer2011Fundamental}

\subsection{Simulating \ce{^{11}B(p,\alpha) 2\alpha} Reaction}

\begin{figure}[hbt]
  \includegraphics[width=1.0\textwidth]{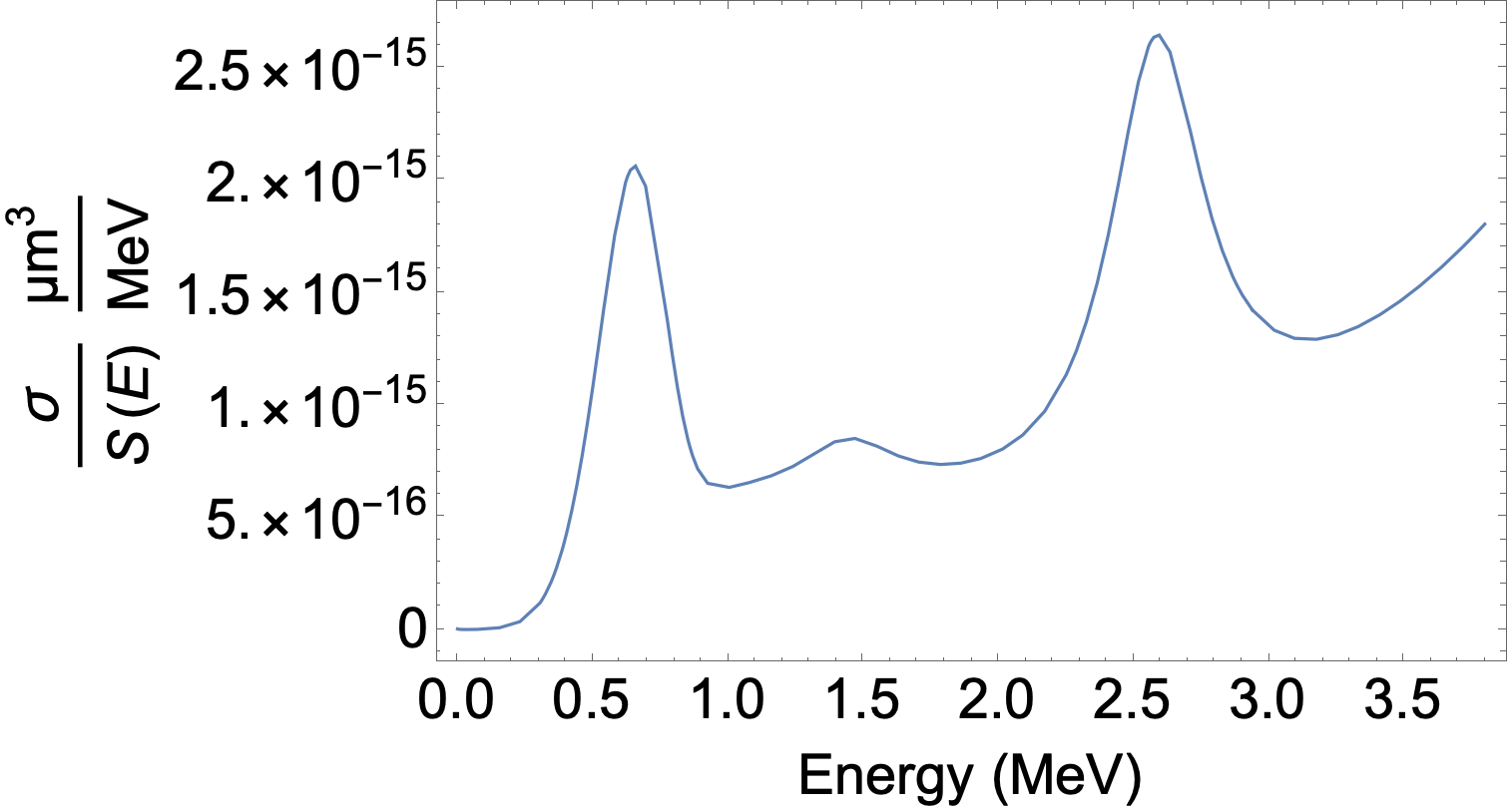}
  \caption{Total cross section/stopping power ratio for the \ce{^{11}B(p,\alpha)^{8}Be} reaction used as the integrand in Equation \ref{eq:Brectionfraction} \cite{zieglerSRIMStoppingRange2008, sikoraNewEvaluation11B2016}.  This was applied to calculations of the reaction fraction for combinations of incident and exiting proton energies.}
  \label{fig:CrossSectSPvsE}
\end{figure}

\begin{figure}[hbt]
  \includegraphics[width=1.0\textwidth]{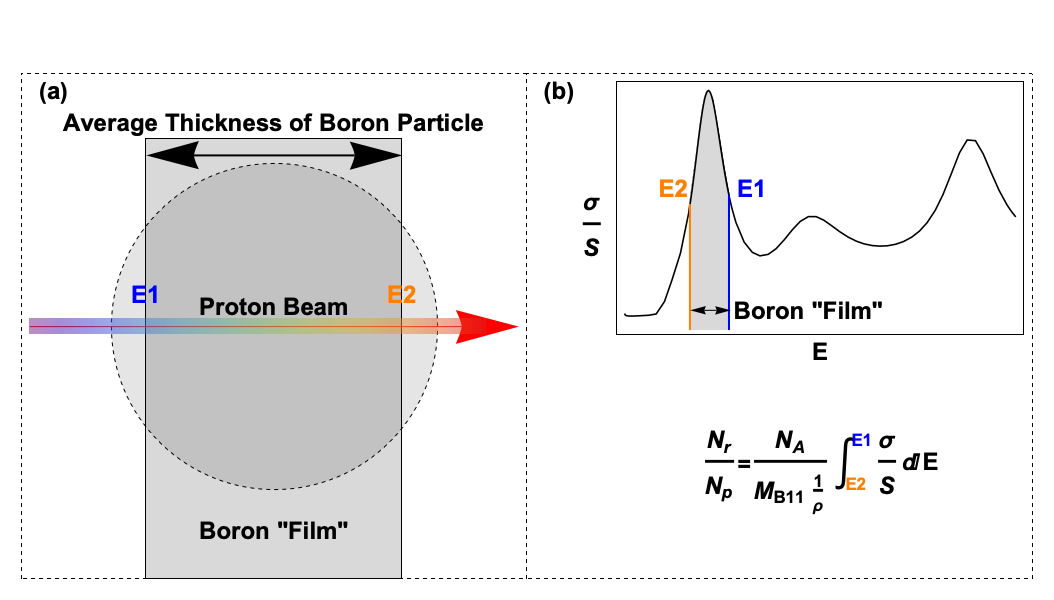}
  \caption{Calculating reaction fractions for a boron particle: (a) A boron particle size is selected (represented by the dashed circle) and the average thickness of the particle is calculated based on its diameter.  Using the thickness and assuming an incident proton energy ($E_1$), the exiting energy ($E_2$) is determined from the energy loss of a proton traversing a boron `film' of that thickness. (b) The number of reactions per proton is computed using Equation \ref{eq:Brectionfraction} with the limits from (a) as $E_2$ is varied and the process is repeated to maximize the number of reactions per proton, which is represented by the shaded area under the curve.}
  \label{fig:reactiongen}
\end{figure}

Creating an accurate simulation of the reaction requires taking into account every proton's energy loss as it traverses a particle, every possible direction each alpha particle could travel, and creating enough reactions to be statistically significant.  The purpose of this simulation was to compare the alpha-energy products of different-sized boron particles and in so doing used several approximation techniques.  The average thickness of a given boron particle along the proton beam direction, $\frac{4}{3}r$ where $r$ is the particle radius, was utilized in calculating average energy loss for an incident proton beam traversing the boron particle, thus replacing the spherical particle with an equivalent film of uniform thickness.    Incident proton energies ($E_2$) were varied to produce the greatest number of alphas per proton for a given boron particle radius using Equation \ref{eq:Brectionfraction} below.  The ratio of total cross section to stopping power that was used as the integrand in Equation \ref{eq:Brectionfraction} is shown in Figure \ref{fig:CrossSectSPvsE}.  This energy-optimizing process is illustrated in Figure \ref{fig:reactiongen}.  Exiting proton ($E_1$) energies were determined by the size of boron particle and the protons' average energy loss from traversing the boron particle. $E_1$ and $E_2$ were chosen so that they bracketed the peak in the reaction cross section at $\approx 660$ keV.  Particle radii were varied between \SI{0.001}{\micro\meter} and \SI{45}{\micro\meter}.  Collisions from the protons in the boron film were modeled using single scattering by the film \cite{evansAtomicNucleus1955} with the film being sufficiently thin that multiple scattering need not be considered.  The reaction fractions (reactions per proton, $\frac{Nr}{Np}$) were computed using a ratio of the total reaction cross section ($\sigma(E)$) and stopping power ($S(E)$) as well as Avogadro's number ($N_A$), molar mass ($M_{B11}$), and mass density ($\rho$) in the equation:

\begin{equation}
  \frac{N_r}{N_p} = \frac{N_A \rho}{M_{B11}} \int_{E1}^{E2} \frac{\sigma(E)}{S(E)}dE.
  \label{eq:Brectionfraction}
\end{equation}

The reaction fraction will vary for different regions within the boron particle because of changing proton energy. To account for this, the particle was divided into concentric cells with different radii.  The region of interest between two radii within the boron particle for alpha generation is termed a cell. Equation \ref{eq:Brectionfraction} was used in two different ways, first to find $E_1$ and $E_2$ for the entire boron particle, and then to find the reaction fraction for each cell (with different $E_1$ and $E_2$).  The reaction fraction for a sphere of radius $R_1$ subtracted from that for a sphere of radius $R_2$ gives the reaction fraction of the cell between $R_2$ and $R_1$ as shown in the equation:
\begin{equation}
  \frac{N(p,3\alpha)}{N_p}_{R2} - \frac{N(p,3\alpha)}{N_p}_{R1} = \frac{N(p,3\alpha)}{N_p}_{\text{cell}}.
  \label{RFR1R2}
\end{equation}
Our simulation divided each boron particle into 20 cells.

%\begin{figure}[hbt]
%  \includegraphics[width=1.0\textwidth]{Cell.png}
%  \caption{Geometry of reaction generation in a cell.  The cell is a region between two radii for which the reaction fraction is calculated.}
%  \label{fig:Shell}
%\end{figure}

\begin{figure}[hbt]
  \includegraphics[width=1.0\textwidth]{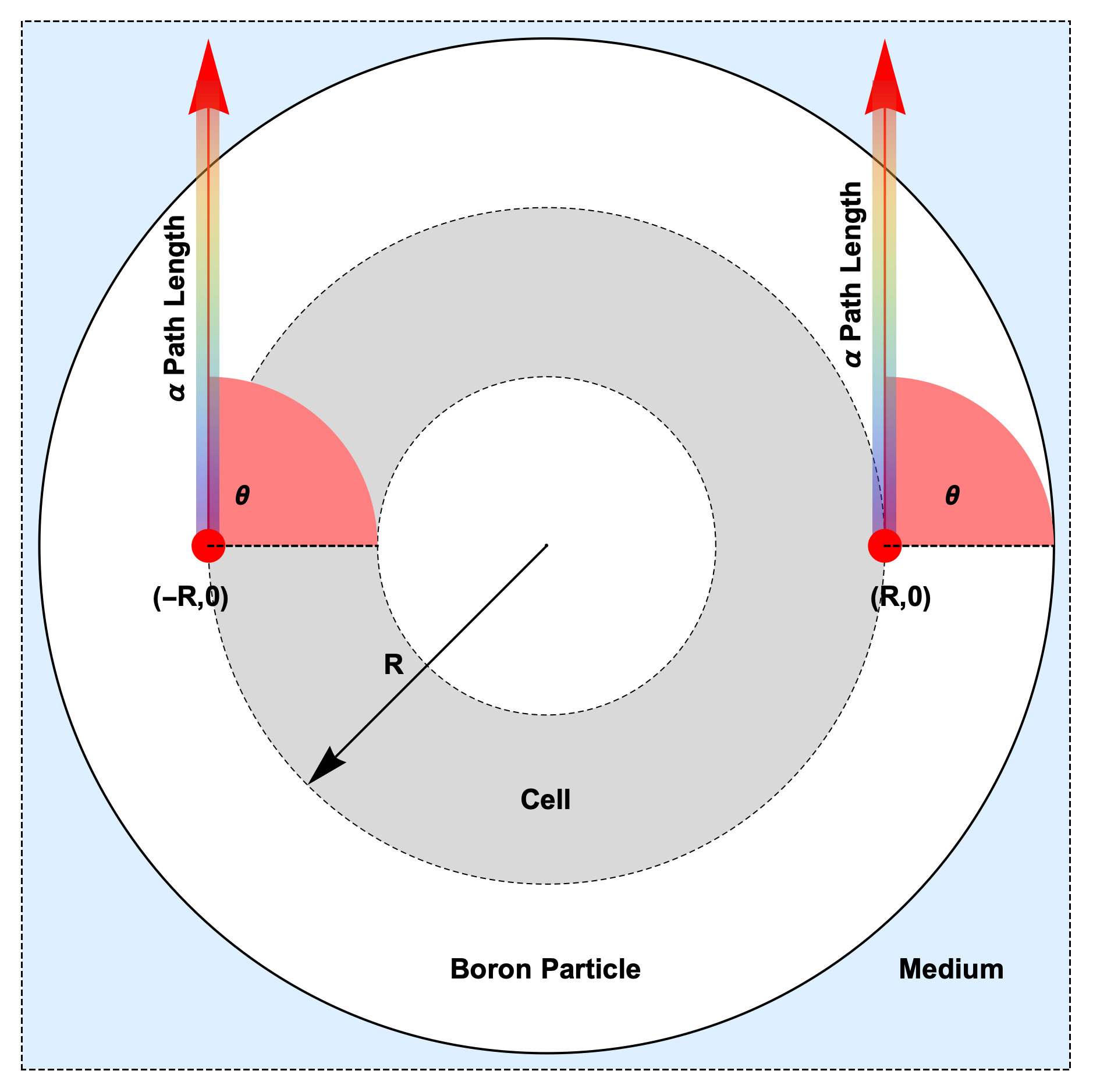}
  \caption{Geometry of energy losses in boron particle in two dimensions.  Alphas are generated at nodes found at (-R,0) and (R,0).  Angles $\theta$ are used to determine the path distance alpha particles would travel from the node to the surface. These calculations are made in the x-y plane. The energy losses from traversing this distance are applied to the generated alphas.}
  \label{fig:alphaescape}
\end{figure}

As mentioned before, the purpose of this simulation was to compare relative energy outputs to the medium surrounding a boron particle.  Due to its broad peak in the alpha energy spectrum as seen in Figure \ref{fig:alphaspectnoE}, $\alpha_1$ was selected as the representative reaction product to compare reactions at given proton energies and angles relative to the proton beam's direction.  The other reaction products are not included, and so the calculation doesn't give total absolute energy deposition in tissue. The effect of the total energy deposition will be approximately twice that calculated for $\alpha_1$ due to the way energy is distributed among the alphas for the reaction. Calculations were carried out for groups of alpha particles emitted in different directions at various depths in the boron particle.  Each \ce{\alpha} group's energy calculations used the average proton incident energy at that depth.  Eight alpha emission angles in the center of mass frame were then selected at equally-spaced intervals from 30$^{\circ}$ to 160$^{\circ}$.  The $\alpha_1$ emission angle and proton energy were used to determine the cross section.  The energy loss of the exiting alpha particles calculation was based on traversed distance from the particle's origin to the surface of the boron particle. The  \ce{^{11}B(p,\alpha)^{8}Be} reactions were simulated at cell points, defined to be at the location of the proton beam's entrance and exit of each cell region on the x-axis as seen in Figure \ref{fig:alphaescape}.

The algorithm itself performs the following steps:\\
1. Select boron particle radius and $\alpha_1$ angles in the center of mass frame and radii by user. Radii were selected from \SI{0.001}{\micro \meter} to \SI{45}{\micro \meter} in \SI{0.090178}{\micro \meter} steps.\\
2. Calculate average thickness of boron particle.  This corresponds to a film.\\
3. Vary and find the incident proton energy $E_1$ and exiting proton energy $E_2$ that produce the maximum number of alpha particles using $\frac{N_r}{N_p}= \frac{N_A \rho}{M_{B11}} \int_{E1}^{E2} \frac{\sigma(E)}{S(E)}$.\\
4. Divide the radius of the current boron particle into uniform intervals to define the radii of 20 concentric spheres.  Convert these concentric spheres into inner films using the average thickness of each concentric sphere.\\
5. Determine the energy of the proton at the entrance and exit of each inner film.  Use these energies to calculate the number of alphas generated at each inner film. \\
6. Calculate the incident and exiting energy of a proton passing through each inner film.  Use these energies and angles in the lab frame to calculate the energy of \ce{\alpha_1} in this direction from \cite{sikoraNewEvaluation11B2016}.\\
7. Calculate the energy lost by alpha particles from the nodes to the surface of the spherical boron particle.\\
8. Add up the energies of escaping alpha particles multiplied by the number of alpha particles generated at each inner sphere and the counts/cross section for each angle in the lab frame for angular weighting.\\
9. Select a new radius for the boron particle and repeat the calculation for the new radius.

%To compare outputs of differing sized particles in a volume the alpha energy product of each radius was multiplied by a volume ratio.  This volume ratio is the quotient of the volume for the maximum particle size investigated, \SI{45}{\micro \meter}, and the volume of the boron particle used to calculate the alpha energy product.  This ratio denotes how many particles could fit in a limited volume of $\frac{4}{3} \pi r_{max}^3$ and is denoted by $\frac{r_{max}^3}{r^3}$.

%A second program was written in Mathematica to simulate and compare equivalent doses for the \ce{^{11}B(p,\ce{\alpha})^{8}Be} reaction.  The number of protons requiring one reaction was also compared.  The reaction fraction for each ``shell" within the boron particle was determined using the same method as the previous simulation.  These reaction fraction were used to model the distribution of alphas in the particle.  Each reaction had a 1\% probability of undergoing the \ce{\alpha_0} channel and a 99\% probability of undergoing the \ce{\alpha_1} channel \cite{kimura2013comment}.

 %The energy of each alpha in the collision was calculated using non-relativistic conservation of momentum and energy.  This included all alphas and decaying Be nuclei.  Beryllium-8 has a very short half life of \ce{t_{1/2}$\thicksim$10^{-16}s} \cite{NuclideChart}.  For the purposes of this simulation it can be approximated that the Beryllium-8 nucleus decays at the point of collision.
 
 \subsection{Simulating Auger Effect}
 
\begin{figure}[hbt]
  \includegraphics[width=1.0\textwidth]{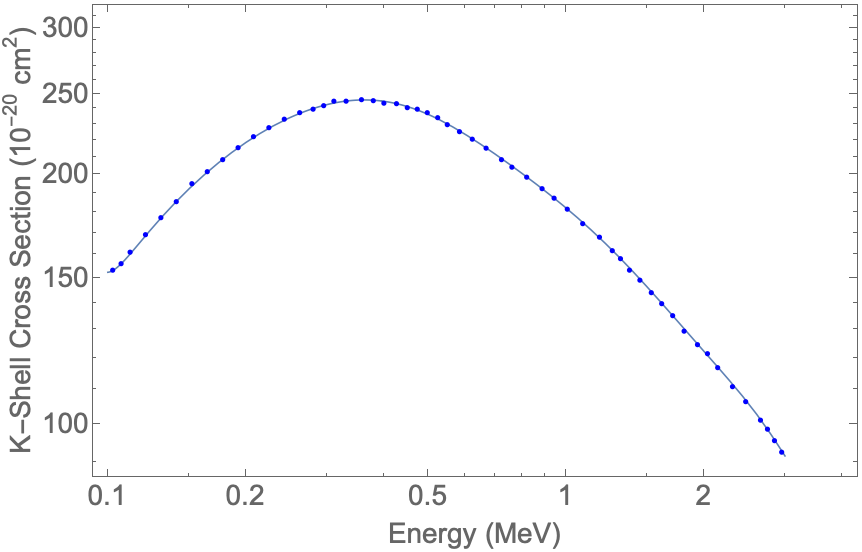}
  \caption{ K-shell ionization cross section in \SI{}{\square \micro\meter} for protons on boron.  Data taken from ECPSSR theoretical predictions \cite{ariyasingheAbsoluteShellIonization1999, brandtShellCoulombIonization1979,brandtEnergylossEffectInnershell1981} and fitted using Mathematica \cite{wolframresearchincMathematica2019}.}
  \label{ECPSSR}
\end{figure}

\begin{figure}[hbt]
  \includegraphics[width=1.0\textwidth]{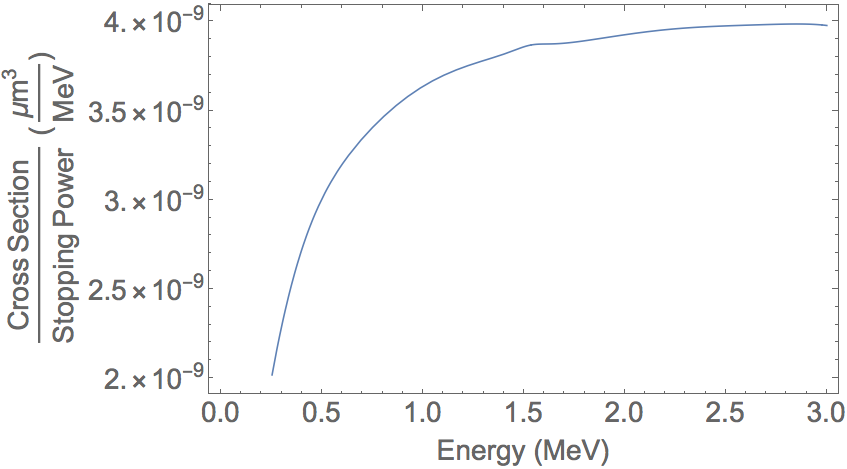}
  \caption{The cross section/stopping power ratio for Auger electrons produced from proton-boron collisions.  This was used to calculate the reaction fraction for combinations of incident and exiting proton energies.  \cite{zieglerSRIMStoppingRange2008, ariyasingheAbsoluteShellIonization1999, brandtShellCoulombIonization1979,brandtEnergylossEffectInnershell1981}}
  \label{fig:AugerCrossStopping}
\end{figure}

A simulation was written to compare electron energy outputs for varying sized boron particles.  This simulates escaping low energy electrons caused by Auger cascades.  Reaction fractions for various cells were calculated in the same way as the \ce{^{11}B(p,\alpha) 2\alpha} reaction simulation, but here using Auger production cross sections.  Incident proton energies were varied from 0-3 MeV.  The Energy-Loss Coulomb-Repulsion Perturbed-Stationary-State Relativistic(ECPSSR) theory \cite{ariyasingheAbsoluteShellIonization1999, brandtShellCoulombIonization1979,brandtEnergylossEffectInnershell1981}, as shown in Figure \ref{ECPSSR}, was used to estimate the K-shell ionization cross sections.  The graphical representation of the K-shell cross section divided by proton stopping power is shown in Figure \ref{fig:AugerCrossStopping}.  All generated electrons were assumed to have a starting energy of 155 eV based on previously published measurements \cite{ariyasingheAbsoluteShellIonization1999}.

\begin{figure}[hbt]
  \includegraphics[width=1.0\textwidth]{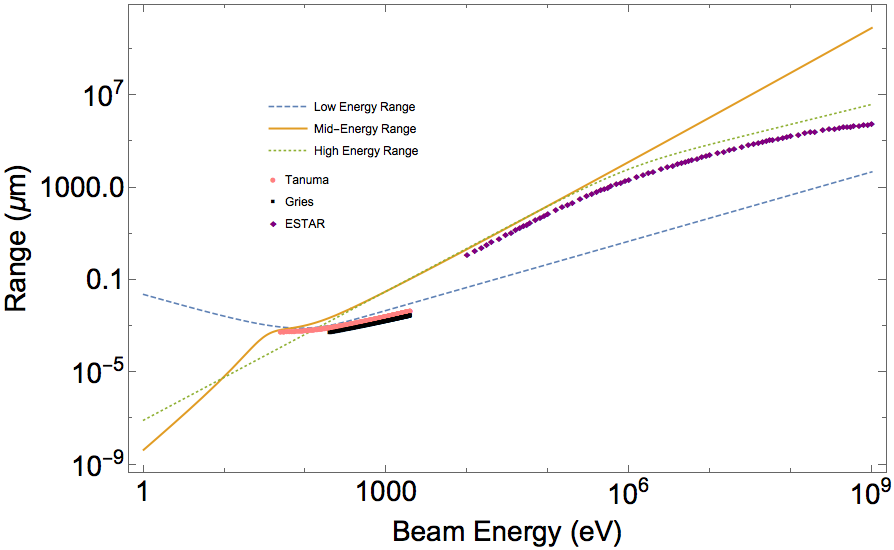}
  \caption{Approximation of range of electrons in boron using a function for different energy ranges \cite{wilsonApproximationRangeMaterials2012}.  Range functions are graphed in comparison with measurements from ESTAR \cite{bergerStoppingPowerRangeTables1998}, Tanuma \cite{tanumaCalculationsElectronInelastic1994}, and Gries\cite{griesUniversalPredictiveEquation1996}.}
  \label{fig:ElectronRange}
\end{figure}

\begin{figure}[hbt]
  \includegraphics[width=1.0\textwidth]{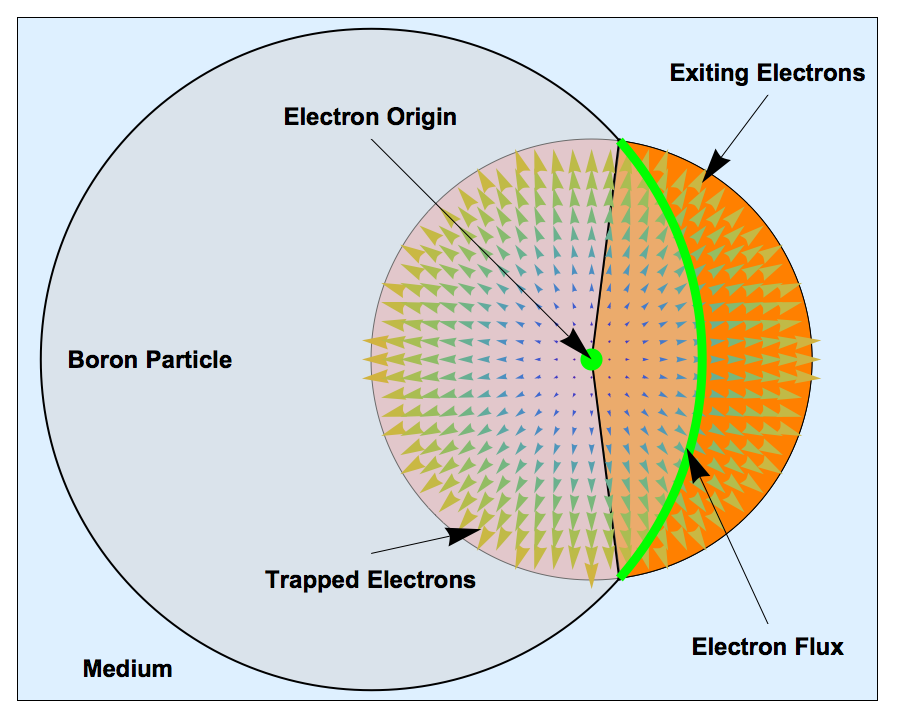}
  \caption{Geometry to find the average path distance from an electron's point of origin to the surface of the boron particle.  The average path distance is only applied to a surface with a non-zero electron flux.}
  \label{fig:ElectronFlux}
\end{figure}

Because all electrons were assumed to have the starting energy of 155 eV this simulation used average Auger path length. This is the average length that electrons would need to travel to escape the boron particle from the node with an energy $>0$. The average path length for each node was then implemented in calculating the energy of escaping electrons. This region of electron flux for the escaping electrons is marked in Figure \ref{fig:ElectronFlux}. Electron range in boron was approximated using the function developed by Wilson and Dennison   \cite{wilsonApproximationRangeMaterials2012}.  The function was found to be suitable for over six orders of magnitude in energy with an uncertainty of $\leq 20\%$ for most conducting, semiconducting, and insulating materials.  The simulation implemented Wilson's mid-energy range approximation for projected range.  Each range of electrons in boron is shown in Figure \ref{fig:ElectronRange}.

\section{Results}

\subsection{\ce{^{11}B(p,\alpha) 2\alpha} Reaction}

\begin{figure}[hbt]
  \includegraphics[width=1.0\textwidth]{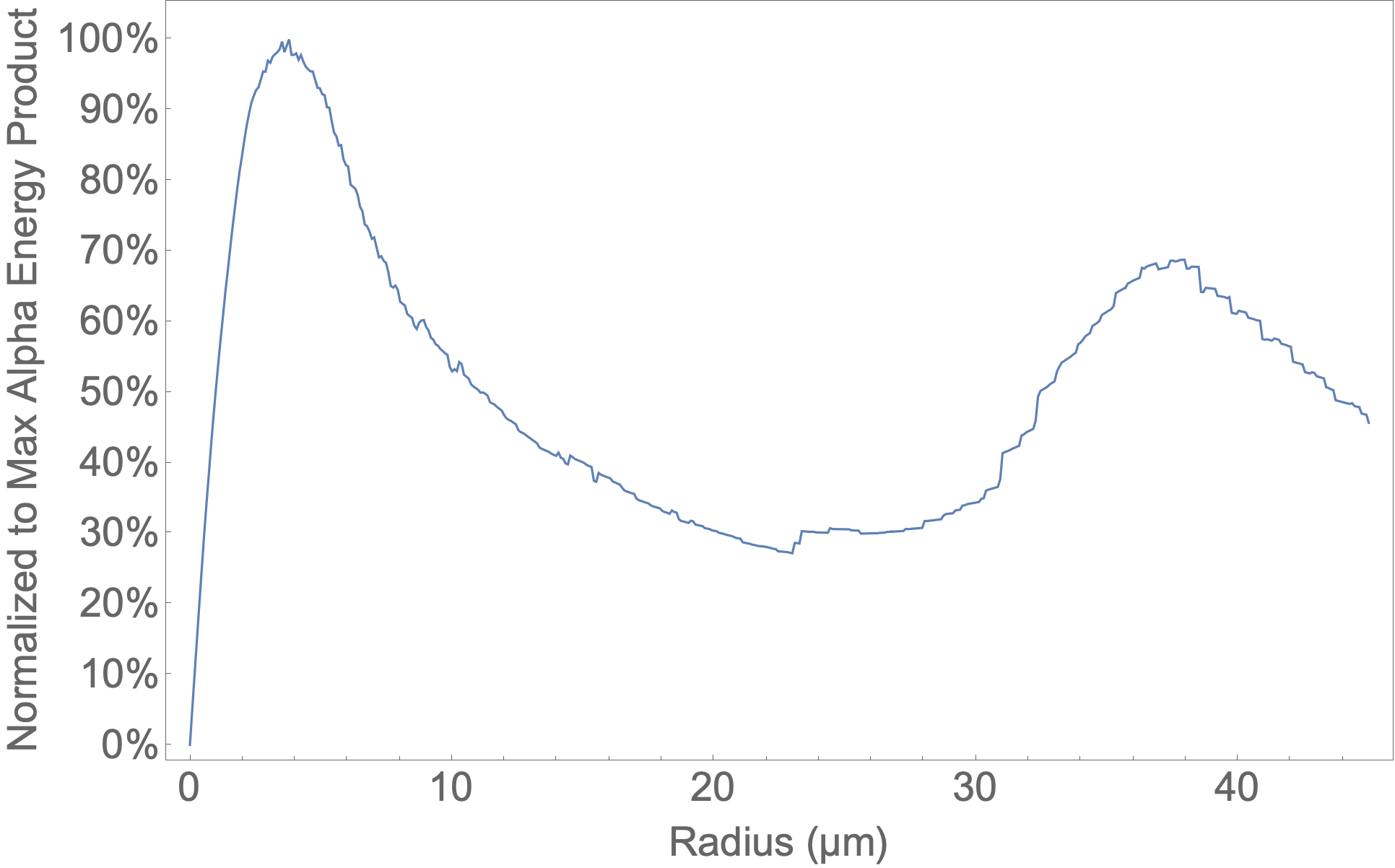}
  \caption{Comparing energy deposited in the surrounding medium for varying sized \ce{^{11}B} particles.  Two notable peaks that produced the highest alpha energy product were found at radius \SI{3.8}{\micro\meter} and \SI{38}{\micro\meter}.}
  \label{fig:B11NPSwo}
\end{figure}

\begin{figure}[hbt]
  \includegraphics[width=1.0\textwidth]{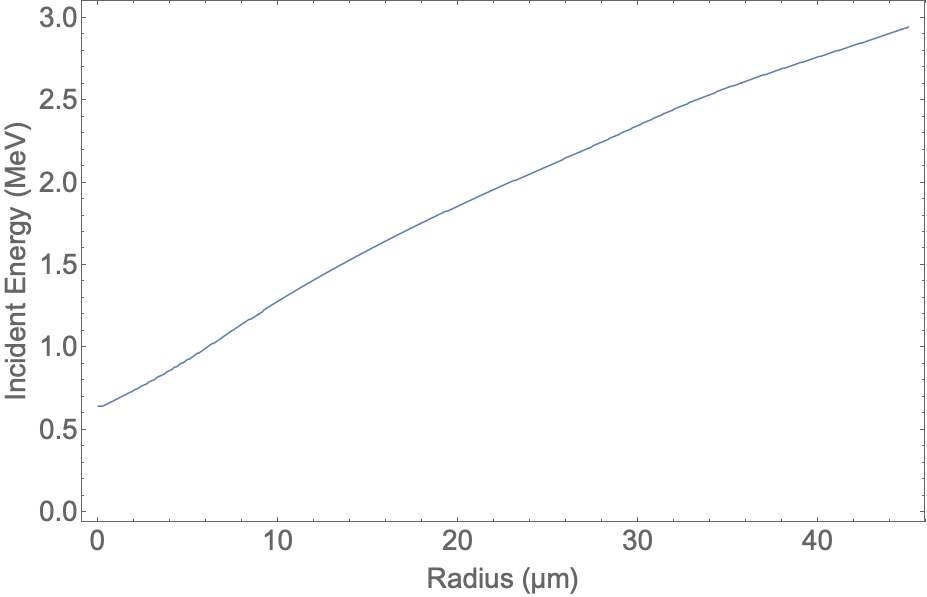}
  \caption{Incident average proton energy for each boron particle that produced the maximum number of alphas.}
  \label{fig:EntRad}
\end{figure}

Figure \ref{fig:B11NPSwo} shows the alpha energy product that exits the boron particle normalized to the maximum output as a percentage vs. the radius.  As the radius of boron increased, so did the alpha energy product until the size of the boron particle began shielding the treatment site, resulting in a maximum alpha energy product at a particle radius of \SI{3.8}{\micro\meter} and again at \SI{38}{\micro\meter}. It would take around 12643  protons to generate an alpha particle for the r = \SI{3.8}{\micro\meter} boron particle and 2960 protons for the r = \SI{38}{\micro\meter} boron particle. These peaks are associated with with the excitation of the two resonances in the reaction at 660 keV and 2.6 MeV.  The average incident proton energy for each boron radius that produces the greatest number of alphas is shown in Figure \ref{fig:EntRad}.  %Figure 16 shows the product of normalized energy output and the volume ratio.  The volume ratio relates how many particles could fit into a volume of $r_{max}$ radius.

\subsection{Auger Effect}

\begin{figure}[hbt]
  \includegraphics[width=1.0\textwidth]{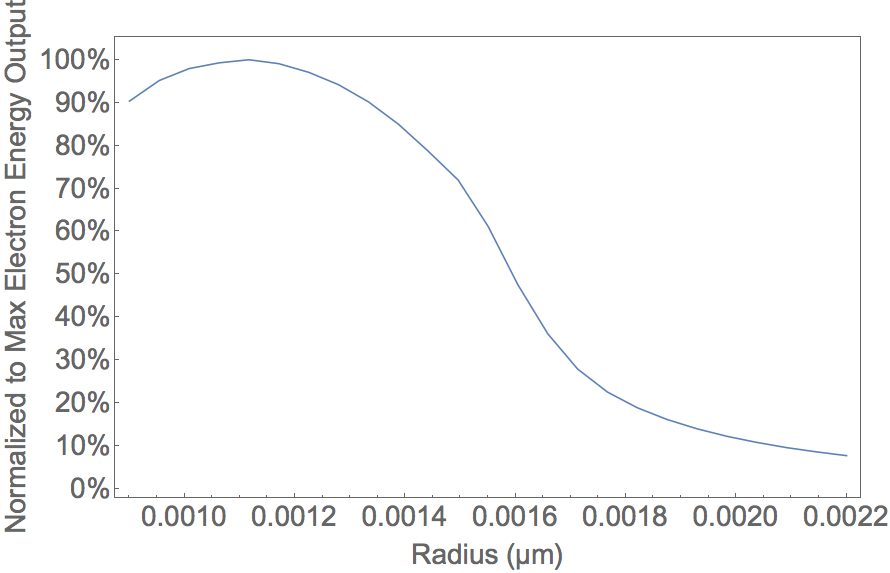}
  \caption{Comparing deposited energy from Auger electrons for varying sized boron particles.  The highest electron energy product size was found to be at radius $\approx$ 1.1 nm.}
  \label{fig:ElectronEnergyProduct}
\end{figure}

Figure \ref{fig:ElectronEnergyProduct} shows the electron energy deposited in the surrounding medium as a percentage of the maximum value.  The radius of boron particle that had the greatest electron energy product was determined to be 1.1 nm.  One reaction would require on average 40 protons for a boron particle at this size.

\section{Conclusions}

The size of the boron particle to maximize alpha dose depends on the limiting range of the biological system.  Energy output can increase with radius until $r=$ \SI{3.8}{\micro\meter}. This size is slightly over the limit for rigid particles passing through the small capillaries: d = \SI{5}{\micro\meter} \cite{baeTargetedDrugDelivery2011}. 

The second peak, at \SI{38}{\micro\meter}, occurs over the particle radius size requirement for embolization therapy: \SI{20}{\micro\meter} \cite{bastianChemoembolizationExperimentalLiver1998}.  This boron particle size could be investigated for use in combination with embolization therapy.  While the alpha energy product may not be maximized at this size, blood restriction may contribute to increasing the treatments efficacy for certain types of liver cancer.

%Larger particles with a diameter of \SI{20}{\micro\meter} have been used for deep lung drug delivery \cite{gonsalves2008biomedical}. Larger particles have been used in embolization therapy and were found in the capillary system of a tumor \cite{bastian1998chemo}.  The particle is also well over the ideal nanoparticle size for brain delivery: 100 nm \cite{olivier2005drug}.  Most tumors have a vascular pore cutoff size range between 380 and 780 nm \cite{hobbs1998regulation}.

PBFT is a promising new treatment method that may one day be implemented to increase the efficacy of proton therapy \cite{blahaProtonBoronReactionIncreases2021, mazzucconiExperimentalInvestigationCATANA2021, cirroneFirstExperimentalProof2018}.  This study only investigated the boron particle that will produce maximum energy output.  Our results will need to be verified experimentally and clinically evaluated on the effect of increased absorbed dose of the surrounding target site.

%Results show the alpha particles would not substantially contribute to the dose of the target site.  While the \ce{^11B(p,\alpha)\ce{2\alpha}} reaction is exothermic, has a large cross section, and produces particles with a greater radiation weighting factor, radiotherapy is severely limited by the amount of protons required to produce one reaction.  The boron particle would likely shield the tumor site from radiation therapy.  Approximately half the proton's energy was distributed within the boron particle instead of the surrounding tissue as shown in Figure 7.  The particle would also require a large amount of protons to raise the temperature and would likely not be suitable for hypertherapy.

The boron particle size to maximize the electron energy product, radius 1.1 nm, is under the size of abundant small pores present in normal tissue endothelium \cite{baeTargetedDrugDelivery2011}.  The energy of the electrons limit their range to 1.12-1.17 nm around the particle \cite{shinotsukaCalculationsElectronInelastic2017}.  %With its larger cross section and smaller dose-maximizing-particle-size, Auger electrons offer an advantage in radiotherapy compared to alphas created from the \ce{^{11}B(p,\alpha)^{8}Be} reaction.

The average energy of Auger electrons simulated was 155 eV which is higher than those used in \cite{walzleinSimulationsDoseEnhancement2014} which had lower than 100 eV for electrons produced in Au and Pt.  This study limited its investigation to only simulating maximum energy output as it relates to particle size.  Future work could experimentally study the surviving fraction of cells in boron-treated colonies with optimum sized boron particles and proton irradiation.  More biological studies are needed to verify its effectiveness.
%The study found that nanoparticles made of these materials may have the potential to enhance dose from proton irradiation.  The experimental work of \cite{polf2011enhanced} also showed that gold nanoparticles could produce clinically meaningful increase in the RBE during proton treatment.
\clearpage

%\section*{References}

%\bibliography{elsarticle-template.bbl}

\end{document}